\documentstyle[12pt,aasms4,epsf]{article}

\begin{document}

\title{Detection of hard X-rays from a Class I protostar\\
 in the HH24-26 region in the Orion Molecular Cloud}
\author{H. Ozawa, F. Nagase, Y. Ueda, T. Dotani, M. Ishida}
\affil{The Institute of Space and Astronautical Science \\
3-1-1, Yoshinodai, Sagamihara, Kanagawa 229-8510, Japan\\
ozawa@astro.isas.ac.jp
}

\begin{abstract}

We observed the HH24-26 region in the L1630 Orion molecular cloud complex
 with the X-ray observatory ASCA in the 0.5$-$10 keV band.
X-ray emission  was detected from the T Tauri star SSV61 and from the region where the Class I protostars
 SSV63E and SSV63W are located (hereafter SSV63E+W). 
The spectra of both SSV63E+W and SSV61 are well explained by an optically thin thermal plasma  model.
The spectrum of the T Tauri star SSV61 has a low temperature of $kT=0.9$ (0.7$-$1.2) keV  and a moderate absorption of $N_{\rm{H}}=1.3$ (0.9$-$1.7) $\times10^{22}$ cm$^{-2}$,
while that of the protostar SSV63E+W has a high temperature of $kT=5.0$ (3.3$-$7.9) keV and  a heavy absorption  of $N_{\rm{H}}=1.5$ (1.2$-$1.8) $\times10^{23}$ cm$^{-2}$. 
The X-ray light curve of SSV63E+W showed a flare
 during the observation.
The peak flux reached  about 9 times that of  the quiescent flux.
The  temperature and the absorption column density do not change 
 conspicuously during the flare. 
The 0.5$-$10 keV luminosity of SSV63E+W was about $1\times10^{32}$ erg s$^{-1}$  in the quiescent state.
The present detection of hard X-rays from SSV63E+W is remarkable, because 
 this is the first X-ray detection of a protostar in Orion.

\end{abstract}


\section{Introduction}

Young stellar objects (YSOs) are divided into 4 classes by the spectral-energy
 distribution at infrared to millimeter wave lengths (Andr\'{e} \& Montmerle 1994):
Class 0 and  Class I protostars,
Class II sources (classical  T Tauri stars, CTTS) and 
Class III sources (weak-line T Tauri stars, WTTS).
The Class 0 protostars have massive envelopes which collapse towards the central region.
Their age  is  estimated to be $\sim$10$^4$ yr. 
The class I protostars have circumsteller disks and dilute dust envelopes.
Their age is estimated to be $\sim$10$^5$ yr.
The Einstein and ROSAT observatories  detected soft X-ray emission from both CTTS and WTTS,
 whose luminosities range from 10$^{28}$ erg s$^{-1}$ to 10$^{30}$ erg s$^{-1}$ 
(Montmerle et al. 1993, Neuh\"{a}user 1997).
Recently, ROSAT and ASCA discovered X-ray 
 emission from Class I protostars (Casanova et al. 1995, Koyama et al. 1996, Grosso et al. 1997, Kamata et al. 1997).
Since X-ray emission has been detected from only a dozen protostars (Carkner et al. 1998),
 it is important to increase the sample size
of X-ray emitting protostars to investigate the origin of their X-ray emission.

The HH24-26 region is located in the L1630 Orion molecular cloud complex at 
 a distance of about 460 pc (Chini et al. 1993).
This region is  an active star-forming region containing several YSOs,
  Class 0 sources (HH24MMS, HH25MMS),  Class I sources (SSV63, SSV59) and 
 a Class II source (SSV61) (Bontemps et al. 1995).
SSV63 and SSV61 were detected in the 2.2 $\mu$ mapping survey (Strom et al. 1976)
 and later SSV63 was resolved into four near-infrared sources, SSV63E, SSV63W, SSV63NE-1, and SSV63NE-2
 (Moneti \& Reipurth 1995).
Both SSV63E and SSV63W show flat or rising spectra in the near- and mid-infrared 
band (Zealy et al. 1992),  and hence belong to Class I.  
SSV63NE-1 and SSV63NE-2 are unclassified diffuse sources,
and may be reflection nebulae (Moneti \& Reipurth 1995).
A complex of Herbig-Haro jets (HH24)  was detected around SSV63
 by optical observations  and some of the jets are probably
 associated with SSV63 itself (Jones et al. 1987, Mundt et al. 1991).
In this paper, we report the first detection of X-ray emission
 from the infrared protostar SSV63.

 \section{Observations}

We observed the HH24-26 region  with the ASCA observatory 
(Tanaka et al. 1994) for a net exposure time of about 30 ks on 1998 October 2.
The center  coordinates of the pointing position were  R.A.(J2000)$=5^{\rm{h}} 46^{\rm{m}} 27.6^{\rm{s}}$ and Decl.(J2000)$=-0^{\circ} 7' 58.7''$.
ASCA has four identical X-ray telescopes (XRT), which achieve
 large effective area at energies up to 10 keV (Serlemitsos et al. 1995).
Two Solid state Imaging Spectrometers (SIS0, SIS1) and
 two Gas Imaging Spectrometers (GIS2, GIS3) are located at each focus
 of an XRT. 
Each SIS has four CCD chips, which have an 11' $\times$ 11' field of view each.
These instruments cover the energy range from 0.4 keV to 10 keV (Burke et al. 1991)
 with an energy resolution of  $\sim$160 eV (FWHM) at 6 keV 
 and $\sim$100 eV (FWHM) at 1.5 keV at the time of  the present observation.
The GIS  has a  field of view of 40' diameter and covers
 the energy range from 0.7 keV to 10 keV 
with an  energy resolution of 0.5 keV (FWHM) at 5.9 keV
(Ohashi et al. 1996).
In this observation, the SIS was operated in the 2-CCD mode, while the GIS
 was in the normal pulse-height  mode with nominal bit assignments.

\section{Results}
\subsection{X-ray image of HH24-26 region}

Figure \ref{image} shows the SIS  soft band (0.7$-$2.0 keV) and hard band (2.0$-$8.0 keV) images
 of the HH24-26 region averaged over the whole observation.
In the figure, the SIS0 and SIS1 images are superimposed
and smoothed by a gaussian function with $\sigma$ of 0.2'.
The symbols with numbers from 1 to 8 in Figure \ref{image} correspond to the sources cited in the caption.
Clearly, we detected two X-ray sources.
The southern source is conspicuous only in the soft band (0.7$-$2.0 keV) image, whereas 
the northern source is predominant in the hard band (2.0$-$8.0 keV) image.
Both detections are consistent with point sources within the angular resolution
 of the XRT (half power diameter about 3').
In order to derive accurate positions and count rates of the detected sources, 
we performed 2-dimensional fitting to the SIS image using the point spread function of the XRT (Ueda 1996).
The positions determined with ASCA are ($5^{\rm{h}}46^{\rm{m}}7.0^{\rm{s}}$, $-0^{\circ}10'7''$) and ($5^{\rm{h}}46^{\rm{m}}7.3^{\rm{s}}$, $-0^{\circ}12'4''$), with 1 $\sigma$ error of 
14''.
Based on the positional coincidence, the southern source is identified with
 the T Tauri star  SSV61.
The northern  source is identified  with the protostars  SSV63E/SSV63W,
 but  the ASCA image resolution does not allow us to resolve X-ray emission
 of the two (hereafter we designate the emission as from SSV63E+W).
The count rates in the 0.7$-$7.0 keV band of SSV61 and SSV63E+W 
are 1.4 (1.3$-$1.6) $\times 10^{-2}$ counts s$^{-1}$ 
(the errors in parentheses indicate a 90\% confidence region)
and 8.2 (7.0$-$9.4) $\times 10^{-3}$ counts s$^{-1}$ 
 , respectively.
We were not able to detect any evidence of X-ray emission from the Class I protostar SSV59 or 
the Class 0 protostars, HH24MMS and HH25MMS.
We derived upper limits (2 $\sigma$) of count rates of these protostars,  2 $\times 10^{-4}$ counts s$^{-1}$, 8 $\times 10^{-4}$ counts s$^{-1}$, and 5 $\times 10^{-4}$ counts s$^{-1}$, respectively for SSV59, HH24MMS, and HH25MMS.

\subsection{Light curve}

Figure 2 shows background-subtracted light curves in the hard (3.0$-$8.0 keV)
 and the soft (0.7$-$2 keV) X-ray bands. 
In the figure, all the data from the  four sensors (SIS0, SIS1, GIS2, GIS3) 
are combined to obtain the best statistics.
The data were taken from regions of 6' radius for the GIS and 4' radius
 for the SIS, centered at the position of SSV63E+W.
The circle shown in Figure \ref{image} indicates the region where the SIS light curve was derived. 
We extracted the background which is constructed  
from the whole FOV of each sensor where point sources were excluded, correcting for the positional dependence of the background intensity (Ueda 1996).
Although  the region contains both SSV61 and SSV63E+W,  
the hard band light curve should correspond to that of  SSV63E+W
 and the soft band light curve to that of SSV61,
as expected from the image analysis in section 3.1.
We discarded the data in the 2.0$-$3.0 keV band from the hard band light curve  
 because X-rays both from SSV63E+W and  SSV61 contribute comparably in the 2.0$-$3.0 keV band, 
 as shown in Figure \ref{spec} of section 3.3.

As can be seen from the top panel of Figure \ref{lc}, SSV63E+W showed
 flare-like time variability.
On the other hand, the soft band light curve (bottom panel of Figure \ref{lc})
 showed little intensity  variability,
 suggesting that the T Tauri star SSV61 was relatively stable during the ASCA observation.
The flare from SSV63E+W is characterized by a slow rise followed by an exponential decay.
We fitted the light curve with a model consisting of a linear rise and an exponential decay  
to a  constant level of the form,   
\begin{equation}
F(t)=\cases{\frac{N}{\tau_{r}}(t-t_{p}+\tau_r)+C,     &  $(t < t_{p})$, \cr 
N exp\{\frac{-(t-t_{p})}{\tau_{d}}\}+C,               &  $(t \ge t_{p}).$       }
\end{equation}
This model has five free parameters, flare peak time $t_{p}$, 
rise time $\tau_{r}$, 
a constant level $C$,
the normalization of the exponential function $N$,
and e-folding decay time $\tau_{d}$.
The solid curve in Figure 2 shows the best-fit model.
From the fit, we derived a peak flux ($N+C$) of 0.18 (0.16$-$0.19) cts s$^{-1}$, 
$C$ = 0.02 (0.016$-$0.026)  cts s$^{-1}$, 
$\tau_{r}$ = 1.7 (1.2$-$2.2) $\times 10^{4}$ s, 
and $\tau_d$ = 1.2 (0.9$-$1.5) $\times 10^{4}$ s.

\subsection{Energy spectrum}

The upper and lower panels in Figure \ref{spec} show background-subtracted spectra 
 taken from the  interval A (hereafter termed the ``flare state'')
 and  the interval B (the ``quiescent state'') indicated in Figure \ref{lc}, respectively.
Since the separation between SSV63E+W and SSV61 ($\sim$2') is not 
large enough compared with the PSF of  the XRT of ASCA,
it is difficult to obtain spectra from SSV63E+W and SSV61 without contamination from the other
 even if we extract the spectra in a small region around each source.
Hence, we accumulated photons in the same region as used in section 3.2 to obtain better statistics. 
The background spectrum is also extracted  from the same region as used in section 3.2. 
Since the region includes both the positions of SSV63E+W and SSV61, 
the spectra will contain contributions from both the sources. 
Both of the spectra in Figure \ref{spec} show
a double peak feature, suggesting the existence of two (hard and soft)
components. Combined with the results from the image analysis in
section 3.1, we assume that the soft and the hard components
correspond to SSV61 and SSV63E+W, respectively. In fact, the soft
component does not show a significant change between the two states,
while the hard component is enhanced during the flare state. This confirms
that the hard component in the spectrum really corresponds to the
emission from the Class I protostar SSV63E+W.

In the hard component during the flare state, an emission line feature is
clearly seen at 6.6 (6.5$-$6.7) keV, which is consistent with the
K$\alpha$ line from He-like iron ions. This suggests that the hard
component during the flare state originates from an optically thin thermal
plasma.  On the other hand, the origin of the soft component from
SSV61 is also considered to be thin thermal coronal emission, which is
typical for T Tauri stars (Montmerle et al. 1993, Neuh\"auther 1997).
Accordingly, we fit the combined GIS and SIS spectra with a model
consisting of two thermal components with different absorption column
densities, in which the high and low temperature components represent
the contribution from SSV63E+W and from SSV61, respectively.  
This model of two thermal components was also adopted for the quiescent spectrum,
utilizing the thin thermal spectral
model calculated by Raymond and Smith (1977; hereafter the ``RS model'').
Since the statistics of the spectrum are not good enough to determine
 the abundance of the soft component, we fixed this  at 0.5 solar  for SSV61 because
 the metal abundances of late-type  stars are roughly 0.5 solar (Carkner et al. 1996). 
For the hard component, we used an Auxiliary Response Function (ARF) constructed for a point source
located in the center of the region selected (i.e., SSV63E+W), while 
for the soft component, we used an ARF constructed for a point source located at the position of SSV61.

Using this ``two-component'' RS model, we obtained
acceptable fits for both of the flare-state and quiescent state spectra. 
The temperature and the absorption column density did
not change significantly between the flare state and the quiescent state for
either the hard or soft components. Hence, to make the tightest
constraints on these parameters, we extracted spectra averaged over
the whole observation, and fit them with the same model. We
again obtained acceptable fits with parameters listed in Table \ref{tbl-2}. 
We derived plasma temperatures $kT=$ 0.9 (0.7$-$1.2) and 5.0 (3.3$-$7.9) keV, 
time-averaged emission measures of 5.3 (3.2$-$9.0) $\times$ 10$^{54}$ cm$^{-3}$ 
and 14 (9.5$-$25) $\times$ 10$^{54}$ cm$^{-3}$,
and absorption column densities  
$N_{\rm H} =$1.3 (0.9$-$1.7) $\times$ 10$^{22}$ cm$^{-2}$ and 
 1.5 (1.2$-$1.8) $\times$ 10$^{23}$ cm$^{-2}$, respectively for the low and  high temperature 
 components.
We obtained an abundance of 0.45 (0.28$-$0.69) solar for the high temperature component, SSV63E+W.
Finally, to determine the
spectral parameters of SSV63E+W separately during the flare state and the
quiescent state, we repeated the fit by fixing the
parameters of the soft component at the best-fit values obtained from
the whole observation. The results are also listed in Table
\ref{tbl-2}.

\section{Discussion}

Since X-ray emission from Class 0 protostars has not been detected yet,
 it is interesting  whether Class 0 protostars in the HH24-26 region,
  HH24MMS and HH25MMS,  emit X-rays or not.
The present ASCA observation, however, failed to detect X-ray emission from 
either HH24MMS or HH25MMS.
We estimate upper limits of X-ray luminosity as 
9 $\times$ 10$^{30}$ erg s$^{-1}$ for HH24MMS,
 7 $\times$ 10$^{30}$ erg s$^{-1}$ for HH25MMS, and 
 4 $\times$ 10$^{30}$ erg s$^{-1}$ for the Class I protostar SSV59,
assuming a  RS model with  $kT=$ 5 keV, $N_{\rm{H}}=$ 1 $\times$ 10$^{23}$ cm$^{-2}$, 
and a metal abundance of 0.5 solar.
We note that these estimates depend on the spectral parameters assumed.

The most remarkable result here is the first detection of X-ray emission from a Class I protostar SSV63E+W 
in Orion.
The significant features of  the X-ray emission are : 
(1) a large X-ray luminosity of about 1 $\times$ 10$^{32}$ erg s$^{-1}$ in the 0.5$-$10 keV band
and a high temperature of  $kT$ = 5.2 ( $>$ 2.3)  keV during the quiescent state, 
(2) an X-ray flare with total energy release of $5 \times 10^{36}$ erg with an elevated temperature 
of $kT \sim 6$ keV, and (3) a large absorption column density of  $N_{\rm{H}}=$ 1.1 (1.0$-$1.3) $\times$ 10$^{23}$ cm$^{-2}$.

SSV63E+W has a high temperature plasma even during the quiescent state, 
which is comparable to that during the flare state. 
Such a high temperature is not observed in TTSs during the quiescent state.
A cluster of Class I protostars in the R CrA molecular cloud also showed
 a high temperature of $kT \sim 7$ keV  during the quiescent state (Koyama et al. 1996).
Thus, some of the Class I sources have high temperature plasmas of $\sim$ 6 keV.
In order to know the mechanism of the quiescent X-ray emission,
we compare the L$_{X}$/L$_{bol}$ ratio of SSV63E+W with those of T Tauri stars.
The bolometric luminosity of SSV63E is estimated to be 22.4 L$_{\odot}$ (Berrilli et al. 1989, Zealey et al. 1992).
Although that of SSV63W is highly uncertain,
assuming  the X-ray emission of SSV63E+W comes from only SSV63E, we obtained L$_{X}$/L$_{bol}$ ratio of
 1.2 $\times$ 10$^{-3}$ in the quiescent state. 
Since the typical L$_{X}$/L$_{bol}$ ratio of T Tauri stars is 5 $\times$ 10$^{-4}$  and the maximum is 3 $\times$ 10$^{-3}$ (Preibisch et al. 1998), our value of the L$_{X}$/L$_{bol}$ ratio of SSV63E+W is comparable to that of T Tauri stars.
Thus the X-ray emission mechanism from protostars during the quiescent state 
has to produce 
 higher temperature plasmas of $\sim$ 6 keV, but not larger  L$_{X}$/L$_{bol}$ ratios than those of T Tauri stars.
The enhanced disk-magnetosphere interaction (Shu et al. 1997) may provide such high temperature plasmas in protostars.

The hard ($kT \sim 6$ keV) X-ray flare of SSV63E+W can probably be explained by strong magnetic activity, 
by analogy with the X-ray flares of TTSs.
Such a hard X-ray flare was also observed  from a protostar 
in the R Cr A molecular clouds (Koyama et al. 1996).
Hayashi et al. (1996) suggested a model for hard X-ray flares in
protostars, based on a closed magnetic loop connecting the central
star and disk. This model can explain the high temperature of X-ray emitting protostars during the flare state.
From the decay time analysis (van den Oord \& Mewe 1989, Montmerle 1990) of the observed flare, 
we estimated the electron density $n_e$ $\sim 1 \times$ 10 $^{11}$ cm$^{-3}$, 
the volume  of the emission region $V \approx E.M./n_e^2$ $\sim 2 \times 10^{33}$ cm$^{3}$, 
the typical length of the  region $d = V^{\frac{1}{3}}$ $\sim 1 \times 10^{11}$ cm, 
and  the magnetic field strength $B$ $\sim$ 200 gauss for the flare region in SSV63E+W. 
These parameters constrain the model of X-ray flare of the protostar.

\acknowledgments
The authors are grateful to Dr. Hugh Hudson and the referee, 
Dr. Thierry Montmerle for their stimulating discussions.
H.O. achnowledges the award of Research Fellowships of the Japan Society for the
 Promotion of Science for Young Scientists.

\newpage
\begin{deluxetable}{ccccc}
\footnotesize
\tablecaption{Best fit spectral parameters\tablenotemark{a} \hspace{0.13cm}and X-ray luminosity for SSV61 and SSV63E+W\label{tbl-2}}
\tablewidth{17cm}
\tablehead{\colhead{} &\colhead{SSV63E+W} &\colhead{SSV63E+W(flare)\tablenotemark{d}} & \colhead{SSV63E+W(quiescent)\tablenotemark{d}}  & \colhead{SSV61}}

\startdata
$kT$ (keV) &
5.0 (3.3$-$7.9)    & 6.3 (3.9$-$11.6)   & 5.2 ($>$2.3)       & 0.9 (0.7$-$1.2) \nl
$N_{\rm{H}}$ (10$^{22}$ cm$^{-2}$) &
15  (12$-$18)      & 14 (12$-$17)       & 14 (8.7$-$23)      & 1.3 (0.9$-$1.7) \nl
abundance &
0.45 (0.28$-$0.69) & 0.55 (0.33$-$0.96) & 0.34 (0.0$-$2.2) & 0.5 (fixed)     \nl
$E.M. $\tablenotemark{b } \hspace{0.13cm}(10$^{54}$ cm$^{-3}$) &
14 (9.5$-$25)      & 21 (14$-$35)       & 7.4 (3.5$-$2.7)  & 5.3 (3.2$-$9.0) \nl
$L\rm{x}$\tablenotemark{c}  \hspace{0.13cm}(10$^{31}$ erg s$^{-1}$) &
22                 & 35                 & 11               &  6.2            \nl
$\chi_\nu^2$/d.o.f. &
68.5/69            & 54.7/56            & 39.7/37         & 68.5/69         \nl

\enddata
\tablenotetext{a}{Single-parameter 90\% confidence regions are shown in parentheses.}
\tablenotetext{b}{The emission measures were calculated from the best-fit GIS parameters.
 Those from SIS parameters are consistent within statistical errors.}
\tablenotetext{c}{Intrinsic X-ray luminosities in the 0.5$-$10 keV band corrected
 for absorption, and assuming a source distance of 460 pc.}
\tablenotetext{d}{
The parameters in the flare state and the quiescent state were obtained
 by fixing the parameters of the soft components to the best-fit values.
}
\end{deluxetable}

\newpage
\figcaption[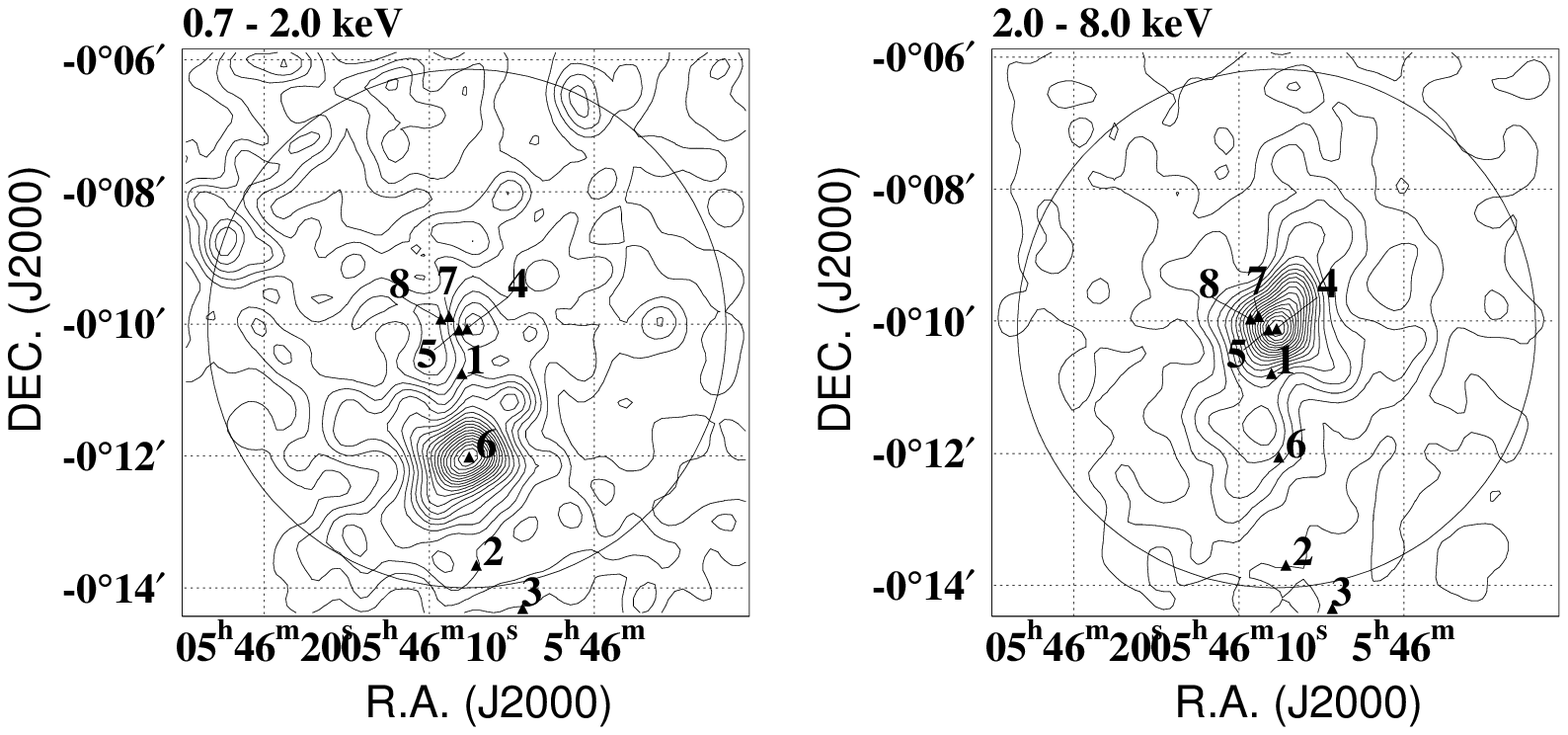]
{The SIS images  of the HH24-26 region in the soft band (0.7$-$2.0 keV,  left panel)
 and in the hard band (2.0$-$8.0 keV, right panel).
The symbol with number 1 corresponds to HH24MMS (Class0), 2 HH25MMS (Class0), 3 SSV59 (Class I), 4 SSV63W (Class I),
 5 SSV63E (Class I), 6 SSV61 (Class II), 7 SSV63NE-1 (reflection nebulae), 8 SSV63NE-2 (reflection nebulae), respectively 
(Cohen \& Schwartz 1983, Bontemps et al. 1995, Zealey et al. 1992., Moneti \& Reipurth 1995).
The circle  with 4' radius indicates 
the region where X-ray photons were extracted to produce light curves and spectra 
(see section 3.2 and 3.3).
 \label{image}}

\figcaption[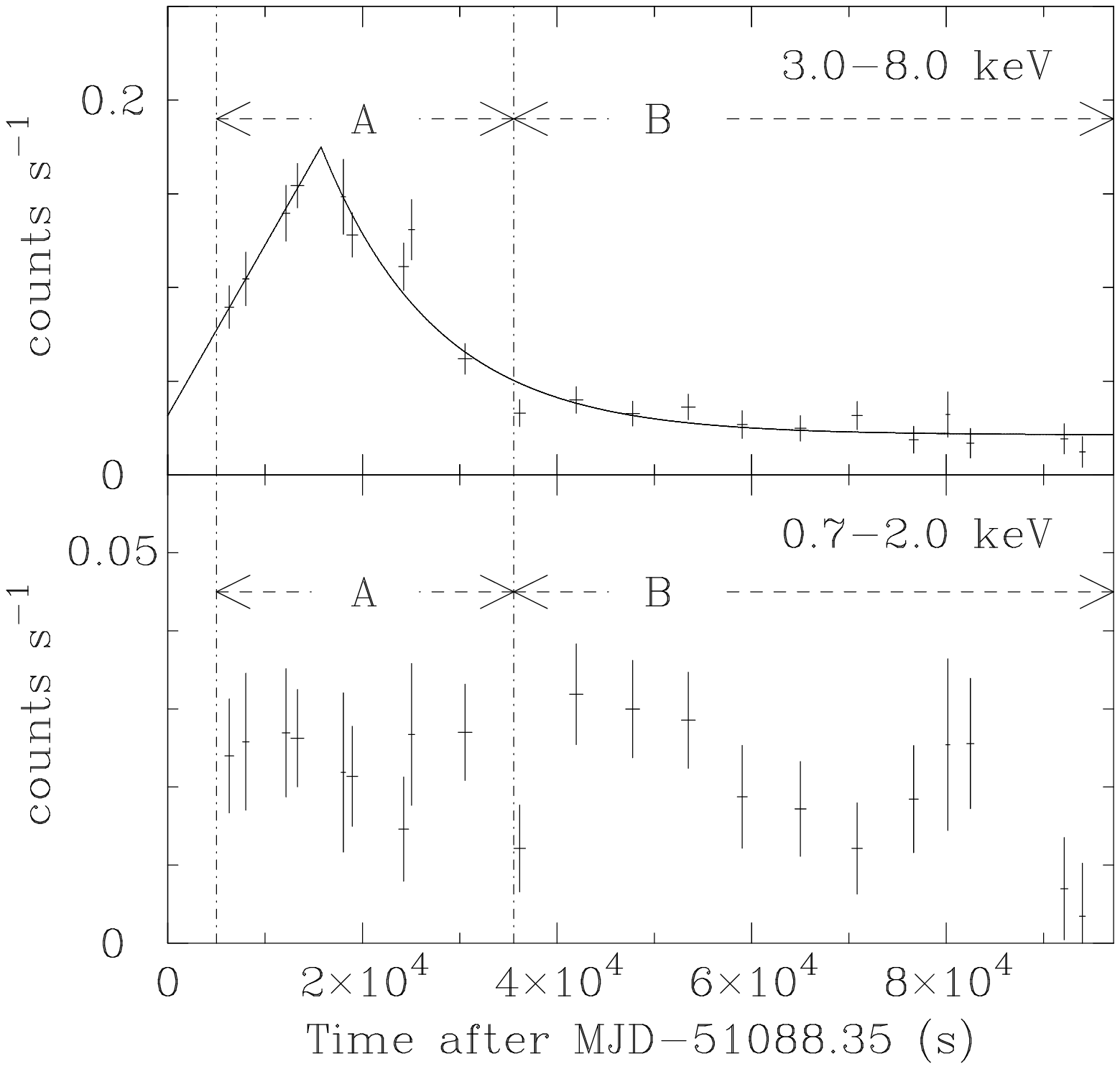]
{X-ray light curves in  the HH24-26 region. 
Upper panel: the light curve in the 3.0$-$8.0 keV band,
 predominantly  corresponds to  the protostar SSV63E+W.
Lower panel: the light curve in the 0.7$-$2.0 keV band,
 predominantly to  the T Tauri star  SSV61. 
The solid line in the upper panel indicates 
the best-fit model (see section 3.2). 
The intervals A and B indicate the flare state and the quiescent state,
 used for extraction of spectra in section 3.3.
\label{lc}}

\figcaption[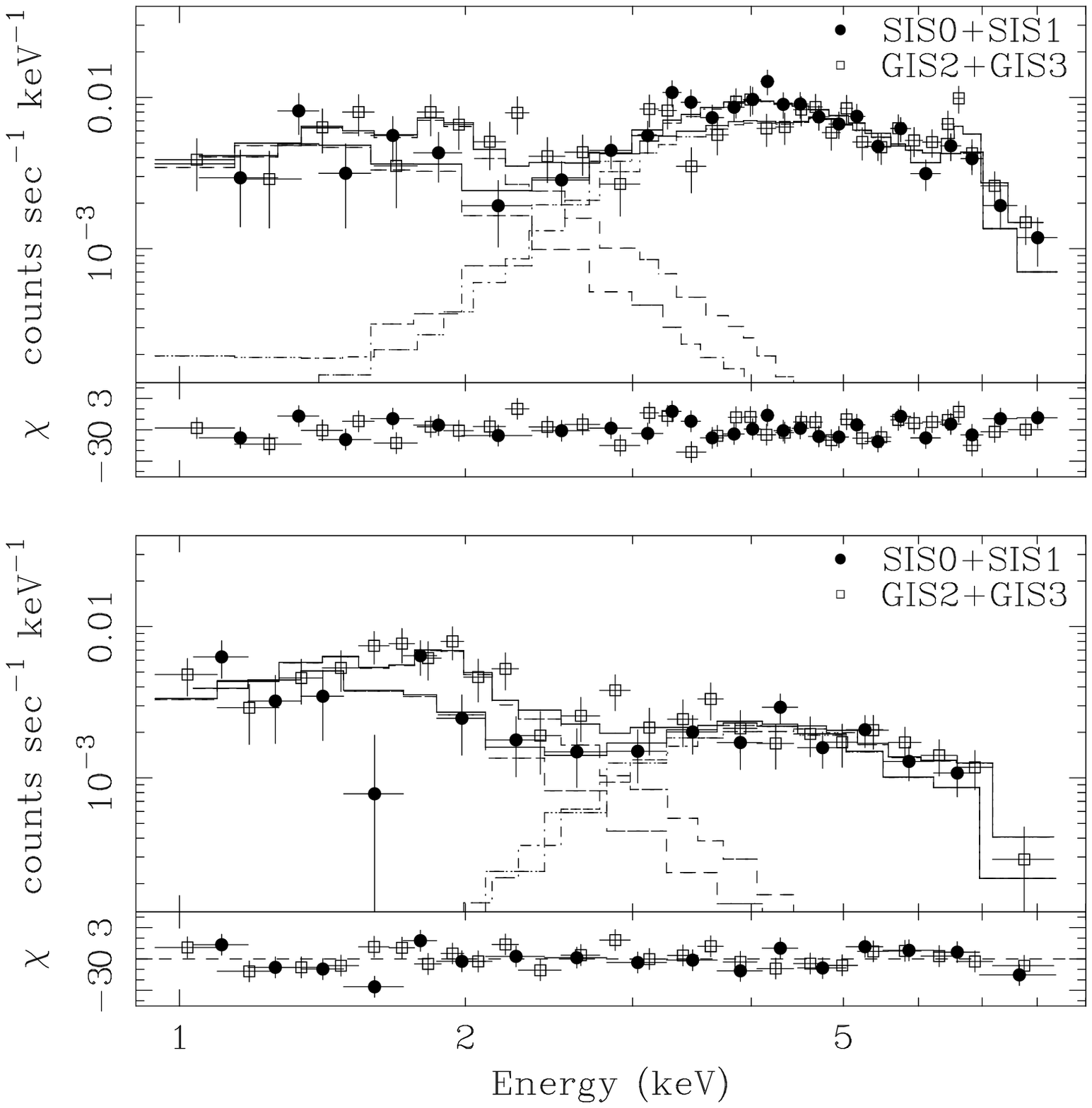]
{GIS (open squares) and SIS (filled circles) energy spectra obtained from the HH24-26 region during the flare state (upper) 
 and during the quiescent state (lower).
The dotted lines indicate the best-fit models of the soft and the hard components,
 and the solid line indicates the total.
The soft component corresponds to the emission from the T Tauri star
 SSV61 and the hard component to that of the protostar SSV63E+W.
 \label{spec}}

\end{document}